\documentstyle[11pt,erice,epsfig]{article}

\def\be{\begin{equation}}
\def\ee{\end{equation}}
\def\bea{\begin{eqnarray}}
\def\eea{\end{eqnarray}}

\begin{document}
\vspace*{4cm}
\title{HEAVY IONS INTERPRETATION OF HIGHEST ENERGY COSMIC RAYS}

\author{STEVE REUCROFT}

\address{Department of Physics, Northeastern University,
\\Boston, MA 02115, USA}

\maketitle\abstracts{A brief review of the energy spectrum of
primary cosmic rays above $10^{10}$ eV and the measurement
techniques used to investigate the ultra high energy ones is
given. This is followed by a discussion of the atmospheric shower
profile of the highest energy event observed using the Fly's Eye
detector. Finally, cosmic ray simulation tools are discussed and
used to investigate the heavy ion interpretation of the highest
energy primary. The best way to contribute further to the
understanding of this issue is by the collection of new and better
data.}

\section{Cosmic Ray Energy Spectrum}

The cosmic ray (CR) spectrum above $10^{10}$ eV (where the Sun's
magnetic field is no longer a concern) can be described by a
series of power laws with the flux falling about 3 orders of
magnitude for each decade increase in energy,\cite{reviews} see
Fig.~1. Above $10^{14}$ eV, the flux becomes so low that direct
measurements using sophisticated equipment on satellites or high
altitude balloons are limited in detector area and in exposure
time. Ground-based experiments with large apertures make such a
low flux observable after a magnification effect in the upper
atmosphere: the incident cosmic radiation interacts with atomic
nuclei of the air molecules and produces extensive air showers
which spread out over large areas. Continuously running using
ingenious installations has raised the maximum observed primary
particle energy to higher than $10^{20}$~eV.\cite{CR}

\begin{figure}
\label{s1}
\begin{center}
\epsfig{file=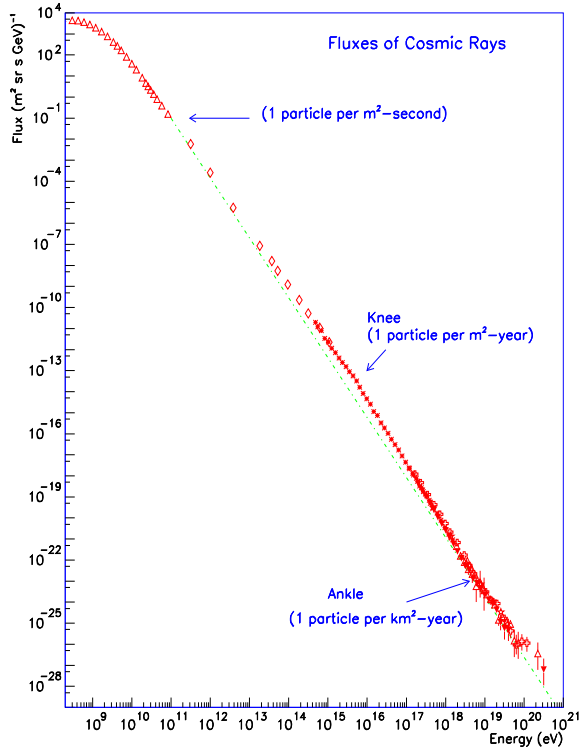,width=7.5cm,clip=}
\caption{Cosmic ray energy spectrum.}
\end{center}
\end{figure}

While theoretical subtleties surrounding CR acceleration provide
ample material for discussion, the debate about the origin of CRs
up to the knee ($\sim 10^{15.5}$ eV) has reached a consensus that
they are produced in supernova explosions.\cite{supernova} The
change of the spectral index (from $-2.7$ to $-3.0$) near the
knee, presumably reflects a change in origin and the takeover of
another, yet unclear type of source. The spectrum steepens further
to $-3.3$ above $\sim 10^{17.7}$~eV (the dip) and then flattens to
an index of $-2.7$ at $\sim 10^{18.5}$~eV (the ankle). A very
widely held interpretation of the modulation features is that
above the ankle a new population of CRs with extragalactic origin
begins to dominate the more steeply falling Galactic
population.\footnote{This hypothesis is supported by AGASA
data.\cite{AGASA}} The origin of the extragalactic channel is
somewhat mysterious.

CRs do not travel unhindered through intergalactic space, as there
are several processes that can degrade the particles' energy. In
particular, the thermal photon background becomes highly blue
shifted for ultrarelativistic protons. The reaction sequence
$p\gamma \rightarrow \Delta^+ \rightarrow \pi^0 p$ effectively
degrades the primary proton energy providing a strong constraint
on the proximity of CR-sources, a phenomenon known  as the
Greisen-Zatsepin-Kuz'min (GZK) cutoff.\cite{gzk} The energy
attenuation length of protons\cite{prd1} is shown in Fig.~2. A
heavy nucleus undergoes photodisintegration in the microwave and
infra-red backgrounds; as a result, iron nuclei do not survive
fragmentation over comparable distances.\cite{nucleus} Ultra high
energy gamma rays would travel even shorter paths due to pair
production on radio photons.\cite{gamma} Therefore, if the CR
sources are all at cosmological distances, the observed spectrum
must virtually end with the GZK cutoff at $E \approx 8 \times
10^{19}$ eV. The spectral cutoff is less sharp for nearby sources
(within 50 Mpc or so).\cite{nearby} The arrival directions of the
trans-GZK events are distributed widely over the sky, with no
plausible optical counterparts (such as sources in the Galactic
Plane or in the Local Supercluster). Furthermore, the data are
consistent with an isotropic distribution of sources in sharp
constrast to the anisotropic distribution of light within 50
Mpc.\cite{hillas}

The difficulties encountered by
conventional acceleration mechanisms in accelerating particles to
the highest observed energies have motivated suggestions
that the underlying production mechanism could be
of non-acceleration nature. Namely, charged and neutral
primaries, mainly light mesons (pions) together with a small
fraction (3\%) of nucleons, might be produced at extremely high
energy by decay of supermassive elementary $X$
particles ($m_X \sim 10^{22} - 10^{28}$ eV).\cite{top-down}
However, if this were the case, the observed spectrum should be
dominated by gamma rays and neutrinos, in contrast to current
observation!~\cite{ave} Alternative explanations involve
undiscovered neutral hadrons with masses above a few GeV,\cite{CFK}
neutrinos producing nucleons and photons via resonant
$Z$-production with the relic neutrino background,\cite{z} or
else neutrinos attaining cross sections in the millibarn range
above the electroweak scale.\cite{nu} If neutrinos/neutral-hadrons
are primaries, they should point back to their sources, thereby enabling
point source astronomy for the most energetic sources of flux at and
above the GZK energy. However, the current CR-sample does not show
significant angular correlation with powerful high-redshift sources.\cite{FB}

\begin{figure}
\label{s2}
\begin{center}
\epsfig{file=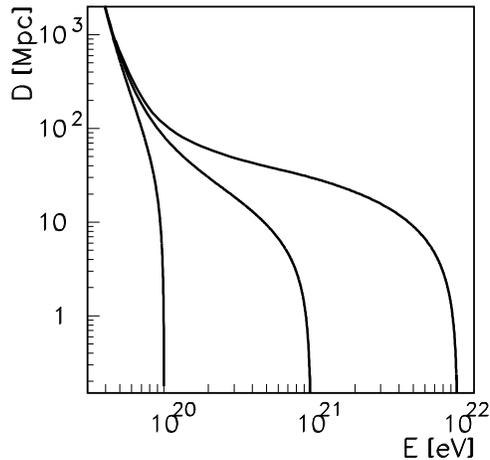,width=7.5cm,clip=} \caption{ Energy
attenuation length of nucleons in the intergalactic medium. Notice
that independently of the initial energy of the nucleon, the mean
energy values approach to 100 EeV after a distance of approx 100
Mpc.}
\end{center}
\end{figure}

\section{Measurement Techniques}

There are two major techniques used in the detection of ultra high energy CRs.
The first one is to build an array of sensors (scintillators, water
Cerenkov tanks, muon detectors) spread over a large area. The detectors
count the particle densities at any given moment, thus sampling the shower
particles hitting the ground. By means of the ground
lateral distributions of the shower one can deduce the direction, the energy and possibly the identity of the primary CR. The Akeno Giant Air Shower
Array (AGASA) is the largest array so far constructed to measure extensive air
showers. This array comprise 111 scintillation detectors each of
2.2 m$^2$ spread over a grid of 100 km$^2$ with 1 km
spacing.\cite{Chiba:1991nf} The array detectors are connected and
controlled through a sophisticated optical
fibre network. The array also contains a number of shielded scintillation
detectors which provide information of the muon content of the showers.
The second technique, pioneered by the University of Utah's Fly's Eye
detector,\cite{NIM} consists in studying the longitudinal
development of extensive air showers by detecting the fluorescence light
produced by the interactions of the charged secondaries. Measurement of
atmospheric fluorescence is possible only on clear, dark nights.
Since mid-1998 a successor to the Fly's Eye instrument, called Hi-Res,
has started taken data at the Dugway site.\cite{sokolsky} In its final
form it is expected to have a time-average aperture  of 340 km$^2$ sr
at $10^{19}$ eV and 1000 km$^2$ at $10^{20}$ eV.

The Auger experiment has been conceived to measure the properties of
cosmic rays above $10^{19}$ eV with unprecedented statistical
precision.\cite{auger}
The Observatory will consist of two sites (in the Northern and Southern
hemispheres) designed to work in a hybrid detection mode, each
covering an area of 3000 km$^2$ with 1600 particle detectors overviewed by 4
fluorescence detectors. Surface array stations are water Cerenkov
detectors (a cylindrical tank of 10 m$^2$ top surface and 1.2 m height,
filled with filtered water and lined with a highly reflective material, the
Cerenkov radiation is detected by 3 photomultiplier-tubes installed at
the top) spaced 1.5 km from each other in an hexagonal grid. These stations
will operate on battery-backed solar power and will communicate with a central
station by using wireless local area network radio links. Event timing will
be provided through global positioning system (GPS) recivers. Of course,
the hybrid reconstruction (see Fig. 3) will be
extremely valuable for energy calibration, but such ``golden'' events
are expected to be less than 10\% of the total event rate.

It is worth mentioning that an array of ground-based detectors is
ideally suited for involving school students and their teachers in
cosmic ray research. This has been proposed several times (for
example, SCROD~\cite{scrod,scrod_2}).

\begin{figure}
\label{s3}
\begin{center}
\epsfig{file=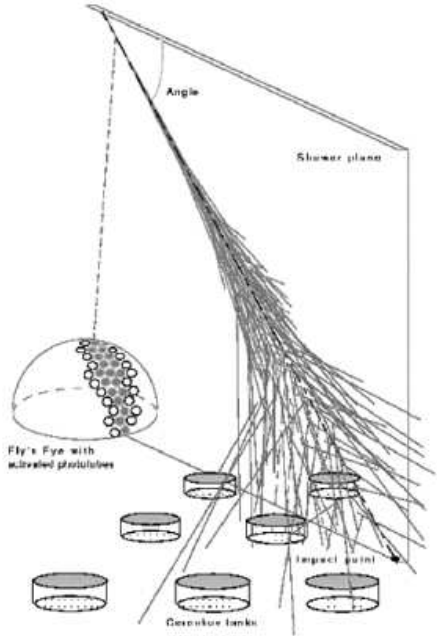,width=7.5cm,clip=}
\caption{Schematic representation of the operation of a hybrid air
shower detector.}
\end{center}
\end{figure}

\section{Shower Profile of Highest Energy Fly's Eye Event}

The Fly's Eye observes an air shower as a nitrogen fluorescence
light source which moves at the speed of light along the path of a
high energy particle traversing the atmosphere.\cite{NIM} In other
words, it directly detects the longitudinal development of the air
cascade. The simulation of the shower evolution depends
sensitively on the first few interactions, necessarily related to
the quality of our ``understandying'' of hadronic collisions. The
well known codes {\sc qgsjet} \cite{qgsjet} and {\sc sibyll}
\cite{sibyll} represent two of the best simulation tools to model
hadronic interactions at the highest energies. The underlying idea
behind  {\sc sibyll} is that the increase in the cross section is
driven by the production of minijets.\cite{sibyll} The probability
distribution for obtaining $N$ jet pairs (with $p_{\rm T}^{\rm
jet}\,>\,p_{\rm T}^{\rm min}$, where $p_{\rm T}^{\rm min}$ is a
sharp threshold on the transverse momentum below which hard
interactions are neglected) in a collision  at energy $\sqrt{s}$
is computed regarding elastic $pp$ or $p\bar{p}$ scattering as a
difractive shadow scattering associated with inelastic processes.
The algorithms are tuned to reproduce the central and
fragmentation regions data up to $p\bar{p}$ collider energies, and
with no further adjustments they are extrapolated several orders
of magnitude. On the other hand, in {\sc qgsjet} the theory is
formulated entirely in terms of Pomeron exchanges. The basic idea
is to replace the soft Pomeron by a so-called ``semihard
Pomeron'', which is defined to be an ordinary soft Pomeron with
the middle piece replaced by a QCD parton ladder. Thus, minijets
will emerge as a part of the ``semihard Pomeron'', which is itself
the controlling mechanism for the whole interaction.\cite{qgsjet}
The different approaches used in both codes to model the
underlying physics show clear differences in cross sections (see
Fig.~4) and multiplicity predictions which increase with rising
energy.\cite{prdhi} As can be seen in Fig.~5, for proton-induced
showers, the differences become washed out as the shower front
gets closer to the ground.

\begin{figure}
\label{s4}
\begin{center}
\epsfig{file=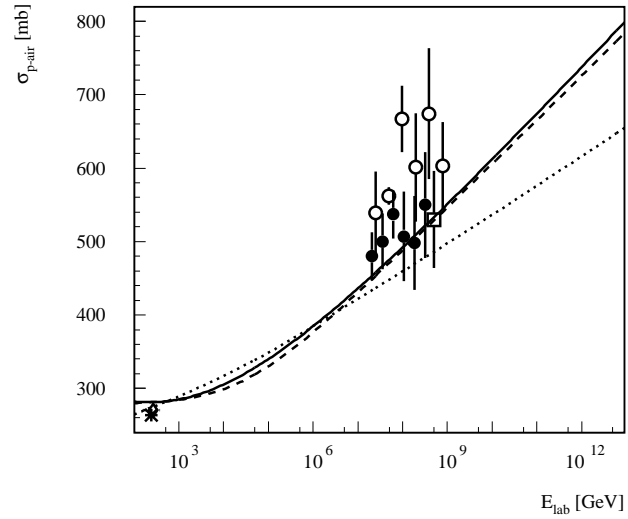,width=8.5cm,clip=}
\end{center}
\caption{$p$-air cross sections of {\sc sibyll} (dashed line), {\sc qgsjet}
(dots), and {\sc aires} (solid line) superimpossed on data obtained from
collider experiments and cosmic ray experiments.~\protect\cite{prdhi}}
\end{figure}

\begin{figure}
\label{s5}
\begin{center}
\epsfig{file=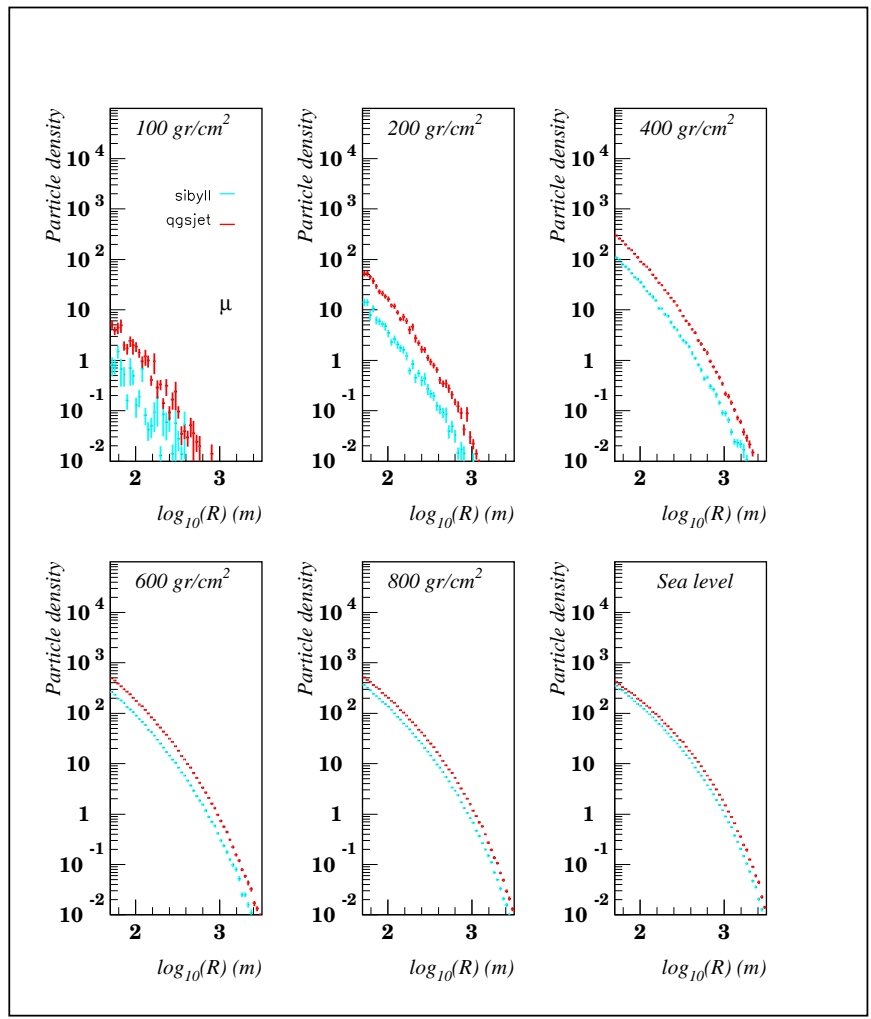,width=11.5cm,clip=}
\caption{Comparison of muon lateral distributions at different
atmospheric depths. $E_p=$ 10 EeV.~\protect\cite{phd}}
\end{center}
\end{figure}

Nevertheless, the footprints of the first hadronic collisions are
still present in the longitudinal development. In Fig.~6 we show
the atmospheric shower profile of 300 EeV showers induced by a
proton, a gold-nucleus and a dust grain\footnote{A very massive
particle containing $10^{7}$ nucleons.\cite{dg}} obtained after
Monte Carlo simulation with the program {\sc aires} (version
2.2.1).\cite{sergio} {\sc sibyll} was used to generate hadronic
interactions above 200 GeV. The simulated results are superimposed
over the experimental data of the highest energy Fly's Eye
event.\cite{FE} It is clearly seen that the gold-nucleus
longitudinal development better reproduces the data than a proton
or a dust grain. However, this is not the case when the hadronic
collisions are modelled with {\sc qgsjet}. As can be seen in
Fig.~7, in this case a medium mass nucleus likely fits the Fly's
Eye data.

\begin{figure}
\label{s6}
\begin{center}
\epsfig{file=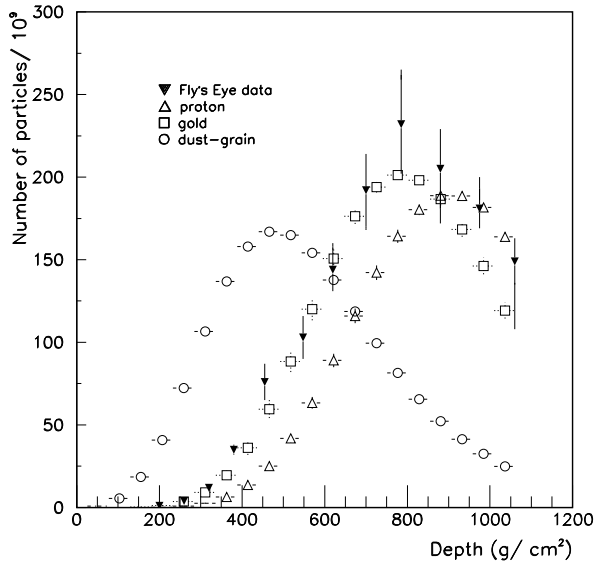,width=7.5cm,clip=}
\caption{Longitudinal development of 300 EeV showers from different
primary species together with the data of the highest energy event
recorded by Fly's Eye.~\protect\cite{golden-s}}
\end{center}
\end{figure}

\begin{figure}
\label{s7}
\begin{center}
\epsfig{file=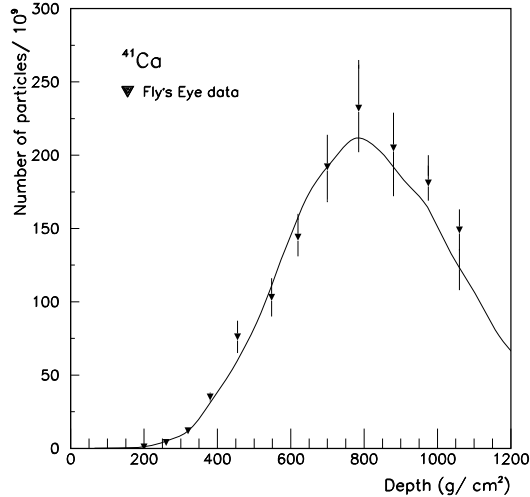,width=7.5cm,clip=}
\caption{Atmospheric cascade development of 300 EeV calcium-nucleus
together with Fly's Eye data.~\protect\cite{neu12}}
\end{center}
\end{figure}

All in all, the primary chemical composition remains hidden by the hadronic
interaction model. In light of this, in the next section we examine in more
detail the speculative case of superheavy nuclei.

\section{Superheavy Nuclei}

It has been generally thought that $^{56}$Fe is a significant end
product of stellar evolution and higher mass nuclei are rare in
the cosmic radiation. Strictly speaking, the atomic abundances of
middle-weight $(60\leq A < 100)$ and heavy-weight ($A>100$)
elements are approximately 3 and 5 orders of magnitude lower,
respectively, than that of the iron group.\cite{BBFH} The
synthesis of the stable super-heavy nuclides  is classically
ascribed to three different stellar mechanisms referred to as the
s-, r-, and p-processes. The s-process results from the production
of neutrons and their capture by pre-existing seed nuclei on time
scales longer than most $\beta$-decay lifetimes. There is
observational evidence that such a kind of process is presently at
work in a variety of chemically peculiar Red Giants \cite{smith}
and in special objects like FG Sagittae \cite{langer} or
SN1987A.\cite{mazzali} The abundance of  well developed nuclides
peaks at mass numbers $A=138$ and $A=208$. The neutron-rich (or
r-nuclides) are synthesized when seed nuclei are subjected to a
very intense neutron flux so that $\beta$-decays near the line of
stability are far too slow to compete with the neutron capture. It
has long been thought that appropriate r-process conditions could
be found in the hot ($T \geq 10^{10}K$) and dense ($\rho \sim
10^{10}-10^{11}$ g/cm$^{3}$) neutron-rich (neutronized) material
located behind the outgoing shock in a type II supernova
event.\cite{delano} Its abundance distribution peaks at $A=130$
and $A=195$. The neutron-deficient (or p-nuclides) are 100-1000
times less abundant than the corresponding more neutron rich
isobars, while their distribution roughly parallels the s- and
r-nuclides abundance curve. It is quite clear that these nuclides
cannot be made by neutron capture processes. It is generally
believed that they are produced from existing seed nuclei of the
s- or r-type by addition of protons (radiative proton captures),
or by removal of neutrons (neutron photodisintegration). The
explosion of the H-rich envelopes of type II supernovae has long
been held responsible for the synthesis of these
nuclides.\cite{BBFH}

Putting all this together, starbursts appear as the natural
sources able to produce relativistic super-heavy nuclei. These
astrophysical environments are supposed to comprise a considerable
population of O and Red Giant stars,\cite{sm} and we believe the
supernovae rate\cite{m-ua-f} is as high as 0.2-0.3 yr$^{-1}$. It
was recently put forward that within this type of galaxies, iron
nuclei can be accelerated to extremely high energies if a two step
process is invoked.\cite{ngc} In a first stage, ions are
diffusively accelerated up to a few PeV at single supernova shock
waves in the nuclear region of the galaxy.\cite{supernova} Since
the cosmic ray outflow is convection dominated, the typical
residence time of the nuclei in the starburst results in $t \sim 1
\times 10^{11}$ s. Thus, the total path traveled is substantially
shorter than the mean free path (which scales as $A^{-2/3}$) of a
super-heavy nucleus. Those which are able to escape from the
central region without suffering catastrophic interactions could
be eventually re-accelerated to superhigh energies at the terminal
shocks of galactic superwinds generated by the starburst. The
mechanism efficiently improves as the charge number $Z$ of the
particle is increased. For this second step in the acceleration
process, the photon field energy density drops to values of the
order of the cosmic background radiation (we are now far from the
starburst region).

The dominant mechanism for energy losses in
the bath of the universal cosmic radiation is the
photodisintegration process. The disintegration rate $R$ (in the system
of reference where the microwave background radiation
is at $2.73$K) of an extremely high energy nucleus with Lorentz
factor $\Gamma$, propagating through an isotropic soft photon
background $n(\epsilon)$ reads,\cite{S}
\begin{equation}
R = \frac{1}{2\Gamma^2} \int_0^\infty d\epsilon\,
\frac{n(\epsilon)} {\epsilon^2}\,\int_0^{2\Gamma\epsilon}
d\epsilon' \,\epsilon' \, \sigma (\epsilon'),
\end{equation}
where $\sigma$ stands for the total photon absortion cross
section. Primed quantities refer to the rest frame of the nucleus.
The total photon absortion cross section is
characterized by a
broad maximum, designated as the giant resonance, located at an energy
of 12-20 MeV depending on the nucleus under consideration. For the medium
and heavy nuclei, $A\geq 50$, the cross section can be well
represented by a single, or in the case of the deformed nuclei, by
the superposition of two Lorentzian curves of the form
\begin{equation}
\sigma(\epsilon') = \sigma_0 \frac{\epsilon'^2 \,
\Gamma_0^2}{(\epsilon^2_0 - \epsilon'^2)^2 \, + \,\epsilon'^2 \,
\Gamma_0^2}\,\,.
\end{equation}
In order to make some estimates, hereafter we refer our
calculations to a gold nucleus (the resonance parameters are
listed in table I).\cite{FW} Figure 8 shows the
energy attenuation length of a gold nucleus due to interactions with the
infrared and microwave backgrounds.\cite{neu2} For comparison, we also show
the energy attenuation length of an iron nucleus. We can conclude that
despite the fact the abundances of superheavy nuclei are around 4 orders of
magnitude lower respective to the iron group, the volume for the potential
sources of extremely high energy superheavy nuclei ($E\approx 300$ EeV)
increases substantially.

\begin{table}
\caption{Giant dipole resonance parameters}
\begin{center}
\begin{tabular}{ccc}
\hline\hline $\epsilon_0$ [MeV] & $\sigma_0$ [mb] & $\Gamma_0$
[MeV]
\\ \hline 13.15 & 255 & 2.9 \\ 13.90 & 365 & 4.0 \\
\hline \hline
\end{tabular}
\end{center}
\end{table}

\begin{figure}
\label{s8}
\begin{center}
\epsfig{file=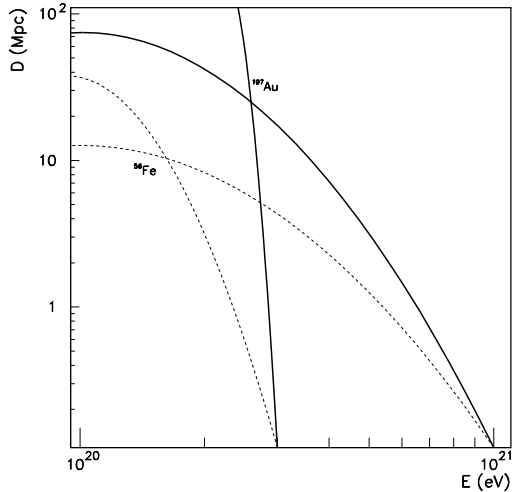,width=7.5cm,clip=}
\caption{Energy attenuation length of gold (solid) and iron (dashed) nuclei
in the intergalactic medium.}
\end{center}
\end{figure}

\section{Conclusions}
The energy spectrum of primary cosmic rays above $10^{10}$ eV has
several compelling features. Perhaps the most impressive is the
apparent occurrence of events above the GZK cutoff. This issue has
been studied using simulation techniques and leads to a plausible
interpretation of the highest energy cosmic ray event as due to
the atmospheric interaction of an energetic heavy ion. The only
reliable way to contribute further to the understanding of this
issue is by the collection of new and better data in future
experiments.

\section*{Acknowledgments}
I acknowledge crucial help from my friend and colleague Luis
Anchordoqui; he provided me with all the figures and most of the
text! I thank Bill Marciano and Sebastian White for organizing such 
an interesting
and eclectic workshop and I am in awe of Antonino Zichichi for
providing Science with such an amazing resource as the one at
Erice. Finally, I am grateful to the US National Science
Foundation for financial support.


\section*{References}
\begin{thebibliography}{99}

\bibitem{reviews} For comprehensive reviews the reader is referred to,
J. Cronin, T. K. Gaisser and S. P. Swordy, Sci. Amer. {\bf 276} 44
(1997); S. Yoshida, and H. Dai, J. Phys. G {\bf 24}, 905 (1998);
M. Nagano and A. A. Watson, Rev. Mod. Phys. {\bf 72}, 689 (2000).

\bibitem{CR} J. Linsley, Phys. Rev. Lett. {\bf 10}, 146 (1963);
M. A. Lawrence, R. J. O. Reid and A. A. Watson,
J. Phys. G {\bf 17}, 773 (1991); M. Takeda {\it et al.}, Phys. Rev. Lett.
{\bf 81}, 1163 (1998); D. J. Bird {\it et al.}, Phys. Rev. Lett. {\bf 71},
3491 (1993); E. Antonov {\it et al.}, JETP Lett. {\bf 69}, 650 (1999).

\bibitem{supernova} P. O. Lagage and C. J. Cesarsky, Astron. Astrophys.
{\bf 118}, 223 (1983); P. L. Biermann, Astron. Astrophys. {\bf 271}, 649
(1993).

\bibitem{AGASA} N. Hayashida et al., Astropart. Phys. {\bf 10}, 303 (1999).

\bibitem{gzk}  K. Greisen , Phys Rev. Lett. {\bf 16}, 748 (1966);
G.T. Zatsepin and V.A. Kuz'min, Pis'ma Zh. Eksp. Teor. Fiz. {\bf
4}, 114 (1966) [JETP Lett. {\bf 4}, 78 (1966)].

\bibitem{prd1} L.A. Anchordoqui, M.T. Dova, L.N. Epele, and J.D. Swain
Phys. Rev. D {\bf 55}, 7356  (1997).

\bibitem{nucleus} J. L. Puget , F. W. Stecker and J. H. Bredekamp, Astrophys.
J. {\bf 205}, 638 (1976). Updated in, L. A. Anchordoqui, M. T. Dova, L. N.
Epele, and J. D. Swain, Phys. Rev. D {\bf 57}, 7103 (1998); L. N. Epele and
E. Roulet, JHEP {\bf 10}, 009 (1998); F. W. Stecker and M. H. Salamon,
Astrophys. J. {\bf 512}, 521 (1999).

\bibitem{gamma} R. J. Protheroe and P. Johnson, Astropart. Phys. {\bf 4},
253 (1996).


\bibitem{nearby} G. R. Farrar, T. Piran, Phys. Rev. Lett. {\bf 84},
3527 (2000); P.L. Biermann, E.J. Ahn, G. Medina Tanco, T. Stanev
Nucl. Phys. B (Proc.Suppl) {\bf 87}, 417 (2000);
L. A. Anchordoqui, H. Goldberg, T. J. Weiler, Phys. Rev. Lett. {\bf 87},
081101 (2001); L. Anchordoqui, H. Goldberg, S. Reucroft, and
J. Swain, Phys. Rev. D {\bf 64},  123004 (2001).

\bibitem{hillas} E. Waxman, K. B. Fisher and T. Piran,
Astrophys. J. {\bf 483}, 1 (1997); M. Hillas, Nature {\bf 395}, 15 (1998).


\bibitem{top-down} P. Bhattacharjee and G. Sigl,
Phys. Rep. {\bf 327} 109 (2000).

\bibitem{ave} M. Ave, J. A. Hinton, R. A. Vazquez and
A. A. Watson, Phys. Rev. Lett. {\bf 85}, 2244 (2000).

\bibitem{CFK} G. R. Farrar, Phys. Rev. Lett. {\bf 76}, 4111 (1996);
D. J. H. Chung, G. R. Farrar and E. W. Kolb, Phys. Rev. D{\bf 57},
4696 (1998).

\bibitem{z} T. J. Weiler, Phys. Rev. Lett. {\bf 49}, 234 (1982);
Astrophys. J {\bf 285}, 495 (1984); Astropart. Phys. {\bf 11}, 303 (1999);
D. Fargion, B. Mele and A. Salis, Astrophys. J. {\bf 517}, 725 (1999).

\bibitem{nu} S. Nussinov and R. Shrock, Phys. Rev. D {\bf 59}, 105002 (1999);
G. Domokos and S. Kovesi-Domokos, Phys. Rev. Lett. {\bf 82}, 1366
(1999); P. Jain, D. W. McKay, S. Panda, and J. P. Ralston, Phys.
Lett. B {\bf 484}, 267 (2000); C. Tyler, A. Olinto and G. Sigl,
Phys. Rev. D {\bf 63}, 055001 (2001); L. Anchordoqui, H.
Goldberg, T. McCauley, T. Paul, S. Reucroft and J. Swain
[hep-ph/0011097]; A. Jain, P. Jain, D. W. McKay and J. P. Ralston
[hep-ph/0011310].

\bibitem{FB} G. R. Farrar, P. L. Biermann, Phys. Rev. Lett. {\bf 81},
3579 (1998); G. Sigl et al., Phys. Rev. D {\bf 63}, 081302 (2001);
A. Virmani, S. Bhattacharya, P. Jain, S. Razzaque,
J. P. Ralston, and D. W. McKay, [astro-ph/0010235].

\bibitem{Chiba:1991nf} N.~Chiba {\it et al.},
Nucl.\ Instrum.\ Meth.\ A {\bf 311}, 338 (1992);
H.~Ohoka, S.~Yoshida and M.~Takeda  [AGASA Collaboration],
Nucl.\ Instrum.\ Meth.\ A {\bf 385}, 268 (1997).

\bibitem{NIM} R. M. Baltrusaitis et al., Nucl. Inst. Methods Phys.
Res., Sect. A {\bf 240}, 410 (1985).

\bibitem{sokolsky} P. Sokolsky, AIP Conference
Proceedings {\bf 433}, 65 (1998).

\bibitem{auger} {\tt http://www.auger.org/admin/}

\bibitem{scrod} {\tt
http://hepnt.physics.neu.edu/scrod/index.html}

\bibitem{scrod_2} L. A. Anchordoqui, J. Cook, M. Gabour, N. Kirsch, 
J. MacLeod, T. P. McCauley, Y. Musienko, T. C. Paul, S. Reucroft, J. D.
Swain, R. Terry, [hep-ex/0106002]. 


\bibitem{qgsjet} N. N. Kalmykov, S. S. Ostapchenko, A. I. Pavlov,
Nucl. Phys. B (Proc. Supp.) {\bf B52}, 17 (1997).

\bibitem{sibyll} R. S. Fletcher, T. K. Gaisser, P. Lipari and T.
Stanev, Phys. Rev. D {\bf 50}, 5710 (1994).


\bibitem{prdhi} L. A. Anchordoqui, M. T. Dova, L. N. Epele and
S. J. Sciutto, Phys. Rev. D {\bf 59}, 094003 (1999).

\bibitem{phd} L. A. Anchordoqui, [astro-ph/9812445].

\bibitem{dg} L. A. Anchordoqui, Phys. Rev. D {\bf 61}, 087302 (2000).

\bibitem{sergio} S. J. Sciutto, in {\it
Proc. XXVI International Cosmic Ray Conference}, (Edts. D. Kieda, M. Salamon,
and B. Dingus, Salt Lake City, Utah, 1999) vol.1, p.411.

\bibitem{golden-s} L. A. Anchordoqui, M. T. Dova, T. P. McCauley, T. Paul,
S. Reucroft, and J. D. Swain, Nucl. Phys.
B (Proc. Suppl.) {\bf 97}, 203 (2001).

\bibitem{FE} D. J. Bird {\it et al.}, Astrophys. J. {\bf 441}, 144 (1995).

\bibitem{neu12} L. A. Anchordoqui, M. Kirasirova, T. P. McCauley,
S. Reucroft, J. D. Swain, Phys. Lett. B {\bf 492}, 237 (2000).

\bibitem{BBFH} E. M. Burbidge, G. R. Burbidge, W. A. Fowler, and F.
Hoyle, Rev. Mod. Phys. {\bf 29}, 547 (1957).

\bibitem{smith} V. V. Smith, AIP Conference Proceedings {\bf 183}, 200 (1989).

\bibitem{langer} G. E. Langer, R. P. Kraft, K. S. Anderson,
Astrophys. J. {\bf 189}, 509 (1974).

\bibitem{mazzali} P. A. Mazzali, L. B. Lucy, K. Butler, Astron.
Astrophys. {\bf 258}, 399 (1992).


\bibitem{delano} M. D. Delano and A. G. W. Cameron, Astrophys.
Space. Sci. {\bf 10}, 203 (1971); W. Hillebrandt, K. Takahashi and
T. Kodama, Astron. Astrophys. {\bf 52}, 63 (1976).

\bibitem{sm} S. Sakai and B. Madore, [astro-ph/9906484].

\bibitem{m-ua-f} See for instance, T. W. B. Muxlow et al., Mont. Not.
Roc. Astron. Soc. {\bf 266}, 455 (1994); J. S. Ulvestad and R. J.
Antonucci, Astrophys. J. {\bf 448}, 621 (1997); D. A. Forbes et
al., Astrophys. J. {\bf 406}, L11 (1993); R. de Grijs et al.,
[astro-ph/9909044].

\bibitem{ngc} L. A. Anchordoqui, G. E. Romero, and J. A. Combi, Phys. Rev.
D {\bf 60}, 10300 (1999).

\bibitem{S} F. W. Stecker, Phys. Rev. D {\bf 180}, 1264 (1969).

\bibitem{FW} E. G. Fuller and M. S. Weiss, Phys. Rev. {\bf 112},
560 (1958).

\bibitem{neu2}L. A. Anchordoqui, M. T. Dova, T. P. McCauley, S. Reucroft,
and  J. D. Swain, Phys. Lett. B {\bf 482}, 343 (2000).



\end{thebibliography}
\end{document}